\documentclass[a4paper,11pt,secnumarabic,amssymb, nobibnotes, prd,footinbib,nofootinbib,ragged2e]{revtex4-1}

\usepackage{geometry}
\usepackage{amsfonts}
\usepackage[pdftex]{graphicx}
\usepackage{multirow}
\usepackage{amsmath}
\usepackage{graphicx}
\usepackage[pdftex]{graphicx}
\usepackage{lipsum}
\usepackage{slashed}
\usepackage{geometry}
\usepackage{multirow}
\usepackage{amsfonts}
\usepackage{placeins}
\usepackage[bottom]{footmisc}
\usepackage[colorlinks=true,linkcolor=red,citecolor=blue]{hyperref}%
\newtheorem{theorem}{Theorem}[section]
\newtheorem{corollary}{Corollary}[theorem]

\begin{document}

\title{Volume and Boundary Face Area of a Regular Tetrahedron in a Constant Curvature Space}%
\author{O.Nemoul}%
\email{omar.nemoul@yahoo.fr}
\address{Laboratoire de physique math�matique et subatomique\\
Mentouri university, Constantine 1, Algeria}

\author{N.Mebarki}%
\email{nnmebarki@yahoo.fr}
\address{Laboratoire de physique math�matique et subatomique\\
Mentouri university, Constantine 1, Algeria}

\date{March 22, 2018}%

\begin{abstract}

An example of the volume and boundary face area of a curved
polyhedron for the case of regular spherical and hyperbolic
tetrahedron is discussed. An exact formula is explicitly derived as
a function of the scalar curvature and the edge length. This work
can be used in loop quantum gravity and Regge calculus in the
context of a non-vanishing cosmological constant.

\end{abstract}

\maketitle
\section{Introduction}

In geometry, the calculation of volume and boundary face area of a
curved polyhedron (geodesic polyhedron\footnote{Geodesic polyhedron
is the convex region enclosed by the intersection of geodesic
surfaces.  A geodesic surface is a surface with vanishing extrinsic
curvature and the intersection of two such surfaces is necessarily a
geodesic curve.}) is one of the most difficult problems. In the case
of spherical and hyperbolic tetrahedra, a lot of efforts has been
made by mathematicians for calculating the volume and boundary face
area: the volume formula are discussed by N. Lobachevsky and L.
Schlafli in refs \cite{r1} for an orthoscheme tetrahedron,  by G.
Martin in ref \cite{r2} for a regular hyperbolic tetrahedron and by
several authors in refs \cite{r3,r4,r5,r6,r7,r8,r9} for an arbitrary
hyperbolic and spherical tetrahedron. All these results are based on
the Schlafli differential equation where a unit sectional curvature
was taken and they are given by a combination of dilogarithmic or
Lobachevsky functions in terms of the dihedral angles. In the
present paper, the volume and boundary face area of a regular
spherical and hyperbolic tetrahedron are explicitly recalculated in
terms of the curvature radius $r=\sqrt{\frac{6}{|R|}}$ and the edge
length $a$. We directly perform the integration over the area and
volume elements to end up with simple formula for the boundary face
area and volume of a regular tetrahedron in a space of a constant
scalar curvature $R$. This can be done by using the projection map
to the Cayley-Klein-Hilbert coordinates system (CKHcs) which maps a
regular geodesic tetrahedron $T(a)$ of an edge length $a$ in the
manifold of a constant curvature $R$ to a regular Euclidean
tetrahedron $T(a_0)$ of an edge length $a_0$ in the CKHcs. Then, one
can express the area and volume measure elements in terms of their
Euclidean ones. A comparison between the regular Euclidean,
spherical and hyperbolic tetrahedron is studied and their
implications are discussed. In physics, a direct application of the
volume and boundary face area of a regular tetrahedron is
essentially in loop quantum gravity (LQG) and Regge calculus. In
LQG, the Euclidean tetrahedron interpretation of a 4-valent
intertwiner state was shown in ref \cite{r10}. The main important
feature of the formula which we are looking for is to find another
possible correspondence between the 4-valent intertwiner state with
a constant curvature regular tetrahedra shapes;  this can be
achieved by inverting the resulted functions. Thus, one can obtain
the scalar curvature measure for a regular tetrahedron shape which
allows us to know what kind of space in which the 4-valent
intertwiner state can be represented by a regular tetrahedron
\cite{r11}. It is worth mentioning that the idea supporting this new
correspondence in the context of LQG with a non-vanishing
cosmological constant was initiated in refs \cite{r11,r12,r13,r14}.
In the context of Regge calculus, the use of a constant curvature
triangulation of spacetime was suggested in ref \cite{r15,r16,r17}
and it can be useful for constructing a quantum gravity version with
a non-vanishing cosmological constant. The paper is organized as
follows: In section \ref{Sec2}, the volume and boundary face area of
a geodesic polyhedron in general curved space are discussed. In
section \ref{Sec3}, we give general integration formula of the
volume and area for constant curvature spaces. In section
\ref{Sec4}, an exact formula for regular spherical and hyperbolic
tetrahedra is explicitly derived as a function of the curvature
radius and the edge length. Finally, in section \ref{Sec5} we draw
our conclusions.

\section{Volume and boundary face area of a polyhedron in a general curved space}\label{Sec2}

For any n-dimensional Riemannian manifold $M$ equipped with an
arbitrary metric $g$ and a coordinates chart $\{U\subset
M,\vec{\tilde{x}}\}$, one has to find another coordinates chart
system $\{U\subset M,\vec{x}\}$, such that the straight lines in the
second are geodesics of the manifold $M$. In other words, it maps
the geodesic curves of the manifold in the first coordinates
system$\ $to the straight line in the second one. Such a coordinates
system denoted by CKHcs (Cayley-Klein-Hilbert coordinates
system)\footnote{It is usually known as the Klein projection.} is
very useful to calculate the volume and boundary face area of a
geodesic polyhedron (i.e. every geodesic polygons and polyhedrons in
the manifold maps to Euclidean polygons and polyhedrons in the CKHcs
respectively). Finding such coordinates system is not an easy task
for general metric spaces because it depends on the geometry itself
and one has to solve a differential equation to find the CKHcs. If
we denote by $\varphi $ the coordinates transformation between the
first and the CKHcs:
\begin{equation}x^A={\varphi }^A(\vec{\tilde{x}})\ \ \ \ \ \ \
\ \ A=\overline{1.n}\,,\label{eq1}\end{equation} one can define the
CKHcs by coordinates transformation that satisfying the following
differential equation (See Appendix \ref{AppA}):
\begin{equation}{\tilde{\nabla}}_V{\tilde{\nabla}}_V{\varphi }^A(\vec{\tilde{x}})=0\,,\label{eq2}\end{equation}
where
\begin{equation}{\tilde{\nabla}}_VV=0\,,\label{eq3}\end{equation}
 Eq. \eqref{eq2} holds for any vector field $V$
tangent to geodesic curves and ${\tilde{\nabla}}_V$ stands for the
covariant directional derivative along the vector field $V$ in the
coordinates system $\{U,\vec{\tilde{x}}\}$. By knowing the metric in
the first coordinates system, one can determine the corresponding
Christoffel symbols $\widetilde{\mathit{\Gamma}}'s$ and then solve
the differential equation \eqref{eq2} to get the ideal frame CKHcs
for calculating the volume of a geodesic polyhedron $Pol$ and its
boundary face area ${\partial Pol}_f$ in an arbitrary n-dimensional
Riemannian space:
\begin{equation}\int_{Pol\subset U\subset M}{{dV}^{Riem}}=\int_{x(Pol)\subset x(U)\subset {\mathbb{R}}^n}{\sqrt{|{det (g(x))\ }|}\ {dV}^{Euc}}\,,\label{eq4}\end{equation}
\begin{equation}\int_{{\partial Pol}_f\subset U\subset M}{{dA}^{Riem}_f}=\int_{x({\partial Pol}_f)\subset x(U)\subset {\mathbb{R}}^n}{\sqrt{{|{det (g(x)|_{{\partial Pol}_f})\ }|}}\ {dA}^{Euc}_f}\,,\label{eq5}\end{equation}
where ${dA}^{Euc}_f$ and ${dV}^{Euc}$ are the Euclidean face area
and volume measures of a geodesic polyhedron respectively, $g(x)$ is
the metric in the CKHcs, $g(x)|_{{\partial Pol}_f}$ is the induced
metric in the geodesic surface ${\partial Pol}_f$.

\FloatBarrier
\begin{figure}[ht]
\begin{center}
\includegraphics[width=2.5in]{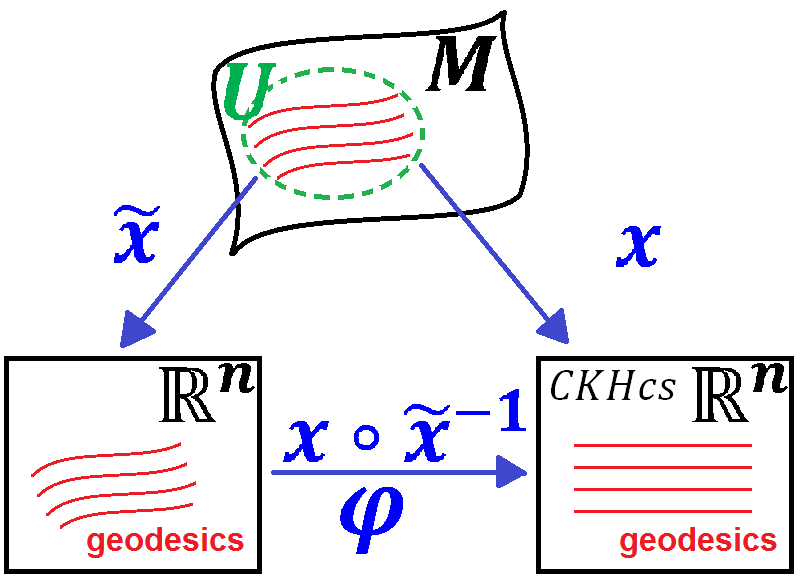}
\end{center}
\caption{The Cayley-Klein-Hilbert coordinates system
(CKHcs).\label{fig1}}
\end{figure}

\section{Volume and boundary face area of a polyhedron in a 3d- constant curvature space}\label{Sec3}

Let $\mathrm{\Sigma }$ be a 3-sphere or 3-hyperbolic metric space.
The metric of the  $S^3_r$ and $H^3_r$ can be combined in a unified
expression and induced from the Euclidean ${Euc}^4$ and the
Minkowski ${Mink}^4$ spaces respectively by using a compact form
$\epsilon $ such that:
\begin{equation}\epsilon =\left\{ \begin{array}{l}
1\ \ \ \ \ \ \ for\ \ \ \ \ \ \ \ \ \ S^3_r\subset {Euc}^4\\
i\ \ \ \ \ \ \ for\ \ \ \ \ \ \ \ \ \ H^3_r\subset
{Mink}^4\end{array} \right.\,,\label{eq6}\end{equation} Let us
consider the cartesian coordinates chart for the two spaces
${Euc}^4$ and ${Mink}^4$
\begin{equation}\begin{array}{l}
X\ \ :\ \ M\ \ \ \ \ \ \ \ \ \ \longrightarrow \ \ \ \ \ \ \ \ \ \ {\mathbb{R}}^3\times \epsilon \mathbb{R}\\
\ \ \ \ \ \ \ \ \ m\ \ \ \ \ \ \ \ \ \ \ \longmapsto \ \ \ \ \ \ \ \
\ \ X^A(m)=\left(x^1,x^2,x^3,\epsilon x^4\right)
 \end{array}
\,,\label{eq7}\end{equation}
 where
\begin{equation}\epsilon \mathbb{R}=\left\{ \begin{array}{l}
\mathbb{R}\ \ \ \ \ \ \ \ \ \ \ \ \ \ \ \ \ \ \ \ \ \ for\ \ \ \ \ \ \ \ \ \ {Euc}^4\  \\
i\mathbb{R}=Im\left(\mathbb{C}\right)\ \ \ \ \ \ \ for\ \ \ \ \ \ \
\ \ \ {Mink}^4 \end{array} \right.\,,\label{eq8}\end{equation}
Basically, the metric of the ${Euc}^4$ and ${Mink}^4$ in this
coordinates system is written as:
\begin{equation}{ds}^2={\delta }_{AB}{dX}^A{dX}^B={\left(dx^1\right)}^2+{\left(dx^2\right)}^2+{\left(dx^3\right)}^2+{\epsilon }^2{\left(dx^4\right)}^2\,,\label{eq9}\end{equation}
In the spherical coordinates $\{\vec{\tilde{x}}\}=\{\rho ,\psi
,\theta ,\varphi \}$ one has:
\begin{equation}\left\{ \begin{array}{l}
\rho =\sqrt{{\delta }_{AB}X^AX^B}\\
\psi =\epsilon arctan\left(\frac{\sqrt{{\left(X^1\right)}^2+{\left(X^2\right)}^2+{\left(X^3\right)}^2}}{X^4}\right) \\
\theta ={arctan \left(\frac{\sqrt{{\left(X^1\right)}^2+{\left(X^2\right)}^2}}{X^3}\right)\ }  \\
\varphi ={arctan \left(\frac{X^2}{X^1}\right)\ }
 \end{array}
\right.\ \ \ \ \ \ \ \ \ \ \ \ \left\{ \begin{array}{l}
X^1=\frac{\rho }{\epsilon }{cos \left(\varphi \right)\ }{sin \left(\theta \right)\ }sin(\epsilon \psi ) \\
X^2=\frac{\rho }{\epsilon }{sin \left(\varphi \right)\ }{sin \left(\theta \right)\ }sin(\epsilon \psi ) \\
X^3=\frac{\rho }{\epsilon }{cos \left(\theta \right)\ }sin(\epsilon \psi )  \\
X^4=\epsilon \ \rho {cos \left(\epsilon \psi \right)\ }
 \end{array}
\right.\,,\label{eq10}\end{equation}

\begin{equation}{ds}^2={\epsilon }^2{d\rho}^2+{\rho }^2\left[{d\psi }^2+{\epsilon }^2{\mathrm{sin}\mathrm{}}^{\mathrm{2}}(\epsilon \psi )\left({d\theta }^2+{\mathrm{sin}\mathrm{}}^{\mathrm{2}}(\theta ){d\varphi }^2\right)\right]\,,\label{eq11}\end{equation}
Now, we define the 3d- metric spaces  $S^3_r$ and  $H^3_r$ as
hyper-surfaces embedded in ${Euc}^4$ and ${Mink}^4$ respectively as:
\begin{equation}X^2={\delta }_{AB}X^AX^B={\left(\epsilon r\right)}^2\,,\label{eq12}\end{equation}
where $r$ is a positive real number known as the radius of
curvature. Geodesics can be obtained by the intersection of $S^3_r$
(or $H^3_r$) surface with two distinct 3d- hypersurfaces through the
centre of the  $S^3_r$ (or $H^3_r$):
\begin{equation}\left\{ \begin{array}{l}
{\delta }_{AB}X^AX^B={\left(\epsilon r\right)}^2 \\
a_AX^A=0 \\
b_AX^A=0 \end{array} \right.\,,\label{eq13}\end{equation} Where
$a_A$ and $b_A$ are two non-collinear vectors of
${\mathbb{R}}^3\times \epsilon \mathbb{R}$. After dividing Eq.
\eqref{eq13} by $cos\left(\epsilon \psi \right)$, the geodesics
satisfy:
\begin{equation}\left\{ \begin{array}{l}
a_1{cos \left(\varphi \right)\ }{sin \left(\theta \right)\ }{tan \left(\epsilon \psi \right)\ }+a_2{sin \left(\varphi \right)\ }{sin \left(\theta \right)\ }{tan \left(\epsilon \psi \right)\ }+a_3{cos \left(\theta \right)\ }{tan \left(\epsilon \psi \right)\ }+a_4=0 \\
b_1\ {cos \left(\varphi \right)\ }{sin \left(\theta \right)\ }{tan
\left(\epsilon \psi \right)\ }+b_2{sin \left(\varphi \right)\ }{sin
\left(\theta \right)\ }{tan \left(\epsilon \psi \right)\ }+b_3{cos
\left(\theta \right)\ }{tan \left(\epsilon \psi \right)\ }+b_4=0
\end{array} \right.\,,\label{eq14}\end{equation}
where $\psi \neq \frac{\pi }{2}$  is used in the case of the
3-sphere $S^3_r$. Therefore, we can get from the geodesic equations
\eqref{eq14}, the coordinates transformation to the CKHcs
$\{\vec{x}\}=\{x,y,z\}$ that satisfying the differential equation
condition \eqref{eq2} for both spherical and hyperbolic cases:

\begin{enumerate}
\item For the spherical case
$S^3_r \ (\epsilon =1\Rightarrow R=\frac{6}{r^2})$ , the coordinates
transformation to the CKHcs and its inverse read:
\begin{equation}\begin{array}{l}
{\varphi }_{S^3_r}:\tilde{x}(U^{S^3_r}\subset S^3_r)\longrightarrow x(U^{S^3_r}\subset S^3_r) \ \ \ \ \ \ \ \ \ \ \ {\varphi }^{-1}_{S^3_r}:x(U^{S^3_r}\subset S^3_r)\longrightarrow \tilde{x}(U^{S^3_r}\subset S^3_r)\\
\ \ \ \ \ \ \ \ \ \ \ \ \ \left(\psi ,\theta ,\varphi
\right)\longmapsto \left(x,y,z\right)\ \ \ \ \ \ \ \ \ \ \ \ \ \ \ \
\ \ \ \ \ \ \ \ \ \ \ \ \ \ \ \ \left(x,y,z\right)\longmapsto
\left(\psi ,\theta ,\varphi \right)
 \end{array}
\,,\label{eq15}\end{equation}
and are defined by
\begin{equation}\left\{ \begin{array}{l}
x=r\ cos\left(\varphi \right)sin\left(\theta \right){tan \left(\psi \right)\ } \\
y=r\ sin\left(\varphi \right)sin\left(\theta \right){tan \left(\psi \right)\ } \\
z=r\ cos\left(\theta \right){tan \left(\psi \right)\ } \end{array}
\right.\ \ \ \ \ \ \ \ \ \ \ \ \ \ \ \ \ \ \ \ \ \ \ \left\{
\begin{array}{l}
\psi =arctan\left(\frac{\sqrt{x^2+y^2+z^2}}{r}\right) \\
\theta =arctan\left(\frac{\sqrt{x^2+y^2}}{z}\right) \\
\varphi =arctan\left(\frac{y}{x}\right) \end{array}
\right.\,,\label{eq16}\end{equation} Notice that  $U^{S^3_r}\subset
S^3_r$ is the top half 3-sphere divided by the hyper-surface of the
equation $\psi =\frac{\pi }{2}$\footnote{Knowing that the biggest
possible spherical tetrahedron is the half of 3-sphere $S^3_r$.}:
\begin{equation}\tilde{x}(U^{S^3_r})=\{(\psi ,\theta ,\varphi)\mathrel{|\vphantom{(\psi ,\theta ,\varphi) \psi \in [0,\frac{\pi}{2}],\theta \in [0,\pi],\varphi \in [0,2\pi ]}.\kern-\nulldelimiterspace}\psi \in [0,\frac{\pi }{2}],\theta \in [0,\pi],\varphi \in [0,2\pi]\}\,,\label{eq17}\end{equation}
\item For the hyperbolic case
$S^3_r \ (\epsilon =i\Rightarrow R=\frac{-6}{r^2})$ , the
coordinates transformation to the CKHcs and its inverse read:
\begin{equation}\begin{array}{l}
{\varphi }_{H^3_r}:\tilde{x}(U^{H^3_r}\subset H^3_r)\longrightarrow {\left[-r,r\right]}^3 \ \ \ \ \ \ \ \ \ \ \ {\varphi }^{-1}_{H^3_r}:{\left[-r,r\right]}^3\longrightarrow \tilde{x}(U^{H^3_r}\subset H^3_r)\\
\ \ \ \ \ \ \ \ \ \ \ \ \ \ \ \left(\psi ,\theta ,\varphi
\right)\longmapsto \left(x,y,z\right)\ \ \ \ \ \ \ \ \ \ \ \ \ \ \ \
\ \ \left(x,y,z\right)\longmapsto \left(\psi ,\theta ,\varphi
\right)
 \end{array}
\,,\label{eq18}\end{equation}
and are defined by
\begin{equation}\left\{ \begin{array}{l}
x=r\ cos\left(\varphi \right)sin\left(\theta \right){tanh \left(\psi \right)\ } \\
y=r\ sin\left(\varphi \right)sin\left(\theta \right){tanh \left(\psi \right)\ } \\
z=r\ cos\left(\theta \right){tanh \left(\psi \right)\ } \end{array}
\right.\ \ \ \ \ \ \ \ \ \ \ \ \ \ \ \ \ \ \ \ \ \ \ \ \ \left\{
\begin{array}{l}
\psi =arctanh\left(\frac{\sqrt{x^2+y^2+z^2}}{r}\right) \\
\theta \ =arctan\left(\frac{\sqrt{x^2+y^2}}{z}\right) \\
\varphi =arctan\left(\frac{y}{x}\right) \end{array}
\right.\,,\label{eq19}\end{equation} Notice that, in order to get an
isomorphism between the two coordinates systems, we have to take the
cubic interval ${\left[-r,r\right]}^3$ since ${\mathrm{tanh}
\left(\psi \right)\ }$ is bounded by the interval
$\left[-1,1\right]$. Moreover, we have also considered the region
$U^{H^3_r}\subset H^3_r$ as the top sheet of the 3d- spherical
hyperboloid $H^3_r$. \end{enumerate} By using the compact form
\eqref{eq6}, one can unify the transformation between the two
coordinates charts for both spherical and hyperbolic cases:
\begin{equation}\left\{ \begin{array}{l}
x=\epsilon r\ cos\left(\varphi \right)sin\left(\theta \right){tan \left(\frac{\psi }{\epsilon }\right)\ } \\
y=\epsilon r\ sin\left(\varphi \right)sin\left(\theta \right){tan \left(\frac{\psi }{\epsilon }\right)\ } \\
z=\epsilon r\ cos\left(\theta \right){tan \left(\frac{\psi
}{\epsilon }\right)\ } \end{array} \right.\ \ \ \ \ \ \ \ \ \ \ \ \
\ \ \ \ \ \ \ \ \ \ \left\{ \begin{array}{l}
\psi =\epsilon \ arctan\left(\frac{\sqrt{x^2+y^2+z^2}}{\epsilon r}\right) \\
\theta =arctan\left(\frac{\sqrt{x^2+y^2}}{z}\right) \\
\varphi =arctan\left(\frac{y}{x}\right) \end{array}
\right.\,,\label{eq20}\end{equation}
The metric in the 3-sphere
$S^3_r$ and 3-hyperbolic $H^3_r$ spaces is:
\begin{equation}{ds}^2=r^2\left[{d\psi }^2+{\epsilon }^2{\mathrm{sin}\mathrm{}}^{\mathrm{2}}(\epsilon \psi )\left({d\theta }^2+{\mathrm{sin}\mathrm{}}^{\mathrm{2}}(\theta ){d\varphi }^2\right)\right]\,,\label{eq21}\end{equation}
Using the differential form chain rule, one can write:
\begin{flalign}& \ \ \ \ \ \ d\psi =\frac{{\epsilon }^2\ r\ x}{({\epsilon }^2r^2+|\vec{x}|^2)|\vec{x}|}dx+\frac{{\epsilon }^2\ r\ y}{({\epsilon }^2r^2+|\vec{x}|^2)|\vec{x}|}dy+\frac{{\epsilon }^2\ r\ z}{({\epsilon }^2r^2+|\vec{x}|^2)|\vec{x}|}dz,&\,\label{eq22}\end{flalign}
\begin{flalign}& \ \ \ \ \ \ d\theta =\frac{x\ z}{|\vec{x}|^2\sqrt{x^2+y^2}}dx+\frac{y\ z}{|\vec{x}|^2\sqrt{x^2+y^2}}dy-\frac{\sqrt{x^2+y^2}}{|\vec{x}|^2}dz,&\,\label{eq23}\end{flalign}
\begin{flalign}& \ \ \ \ \ \ d\varphi =\frac{-y}{x^2+y^2}dx+\frac{x}{x^2+y^2}dy,&\,\label{eq24}\end{flalign}
Thus, the metric in the CKHcs becomes:
\begin{equation}{ds}^2=g_{AB}dx^Adx^B=-{\left(\frac{\sum^3_{A=1}{x^A{dx}^B}}{{\epsilon }^2r^2+{\left|\vec{x}\right|}^2}\right)}^2+\frac{\sum^3_{A=1}{{\left({dx}^A\right)}^2}}{{\epsilon }^2r^2+{\left|\vec{x}\right|}^2}\,,\label{eq25}\end{equation}
The components of the metric elements read:
\begin{equation}g_{AB}=\left( \begin{array}{ccc}
\frac{{\epsilon }^2r^2\ \left({\epsilon }^2r^2+y^2+z^2\right)}{{\left({\epsilon }^2r^2+x^2+y^2+z^2\right)}^2} & \frac{-\ {\epsilon }^2\ r^2\ xy}{{\left({\epsilon }^2r^2+x^2+y^2+z^2\right)}^2} & \frac{-\ {\epsilon }^2\ r^2\ xz}{{\left({\epsilon }^2r^2+x^2+y^2+z^2\right)}^2} \\
\frac{-\ {\epsilon }^2\ r^2\ xy}{{\left({\epsilon }^2r^2+x^2+y^2+z^2\right)}^2} & \frac{{\epsilon }^2r^2\ \left({\epsilon }^2r^2+x^2+z^2\right)}{{\left({\epsilon }^2r^2+x^2+y^2+z^2\right)}^2} & \frac{-\ {\epsilon }^2\ r^2\ yz}{{\left({\epsilon }^2r^2+x^2+y^2+z^2\right)}^2} \\
\frac{-\ {\epsilon }^2\ r^2\ xz}{{\left({\epsilon
}^2r^2+x^2+y^2+z^2\right)}^2} & \frac{-\ {\epsilon }^2\ r^2\
yz}{{\left({\epsilon }^2r^2+x^2+y^2+z^2\right)}^2} & \frac{{\epsilon
}^2r^2\ \left({\epsilon }^2r^2+x^2+y^2\right)}{{\left({\epsilon
}^2r^2+x^2+y^2+z^2\right)}^2} \end{array}
\right)\,,\label{eq26}\end{equation} and the Jacobian $J(\vec{x})$

\begin{equation}J(\vec{x})=\sqrt{\left|{det \left(g(x)\right)\ }\right|}=\frac{r^4\ }{{\left({\epsilon }^2r^2+{\left|\vec{x}\right|}^2\right)}^2}\,,\label{eq27}\end{equation}

 Finally, we can determine the volume of a geodesic polyhedron
$Pol\ $and its boundary face area ${\partial Pol}_f$:
\begin{enumerate}
\item For a spherical polyhedron $(R=\frac{6}{r^2})$
\begin{equation}\int_{{\partial Pol}_f\subset U^{S^3_r}\subset S^3_r}{{dA}^{S^3_r}_f}=\int_{x\left({\partial Pol}_f\right)\subset {\mathbb{R}}^3}{{dA}^{Euc}_f{\sqrt{{|{det (g(x)|_{{\partial Pol}_f}^{S^3_r})\ }|}}}}\,,\label{eq28}\end{equation}
\begin{equation}\int_{Pol\subset U^{S^3_r}\subset S^3_r}{{dV}^{S^3_r}}=\int_{x\left(Pol\right)\subset {\mathbb{R}}^3}{{dV}^{Euc}\frac{r^4\ }{{\left(r^2+{\left|\vec{x}\right|}^2\right)}^2}\ }\,,\label{eq29}\end{equation}
\item For a hyperbolic polyhedron $(R=\frac{-6}{r^2})$
\begin{equation}\int_{{\partial Pol}_f\subset U^{H^3_r}\subset H^3_r}{{dA}^{H^3}}=\int_{x\left({\partial Pol}_f\right)\subset {\mathbb{R}}^3}{{dA}^{Euc}_f{\sqrt{{|{det (g(x)|_{{\partial Pol}_f}^{H^3_r})\ }|}}}}\,,\label{eq30}\end{equation}
\begin{equation}\int_{Pol\subset U^{H^3_r}\subset H^3_r}{{dV}^{H^3_r}}=\int_{x\left(Pol\right)\subset {\mathbb{R}}^3}{{dV}^{Euc}\frac{r^4\ }{{\left(-r^2+{\left|\vec{x}\right|}^2\right)}^2}\ }\,,\label{eq31}\end{equation}
\end{enumerate}
The induced Jacobian $\sqrt{{|{det (g(x)|_{{\partial
Pol}_f}^{S^3_r})\ }|}}$ and $\sqrt{{|{det (g(x)|_{{\partial
Pol}_f}^{H^3_r})\ }|}}$ for both spherical and hyperbolic
respectively can be determined after restricting the metric in the
boundary surface area ${{\partial Pol}_f}$.
\section{Application: Regular tetrahedron in a constant curvature space}\label{Sec4}

Let $T(a)$ be a regular geodesic tetrahedron with an edge length $a$
embedded in a constant curvature 3d- space $\mathrm{\Sigma }$, and
${\left\{{\vec{A}}_f\right\}}_{f=\overline{1.4}}$ be normal areas
vectors of $T(a)$. In what follows, we will calculate the volume of
a geodesic regular tetrahedron $T(a)$ and its boundary face area
${\partial T(a)}_f$ in 3d- sphere $S^3_r$  and Hyperbolic $H^3_r$
manifolds:
\begin{equation}A^{\mathit{\Sigma}}_f(r,a)=\int_{x\left({\partial T(a)}_f\right)\subset {\mathbb{R}}^3}{{dA}^{Euc}_f} \sqrt{\left|det(g(x)|_{{\partial T(a)}_f})\right|} \,,\label{eq32}\end{equation}
\begin{equation}V^{\mathit{\Sigma}}\left(r,a\right)=\int_{x\left(T(a)\right)\subset {\mathbb{R}}^3}{{dV}^{Euc}\frac{r^4\ }{{\left({\epsilon }^2r^2+{\left|\vec{x}\right|}^2\right)}^2}}\,,\label{eq33}\end{equation}

 \FloatBarrier
\begin{figure}[ht]
\begin{center}
\includegraphics[width=2.5in]{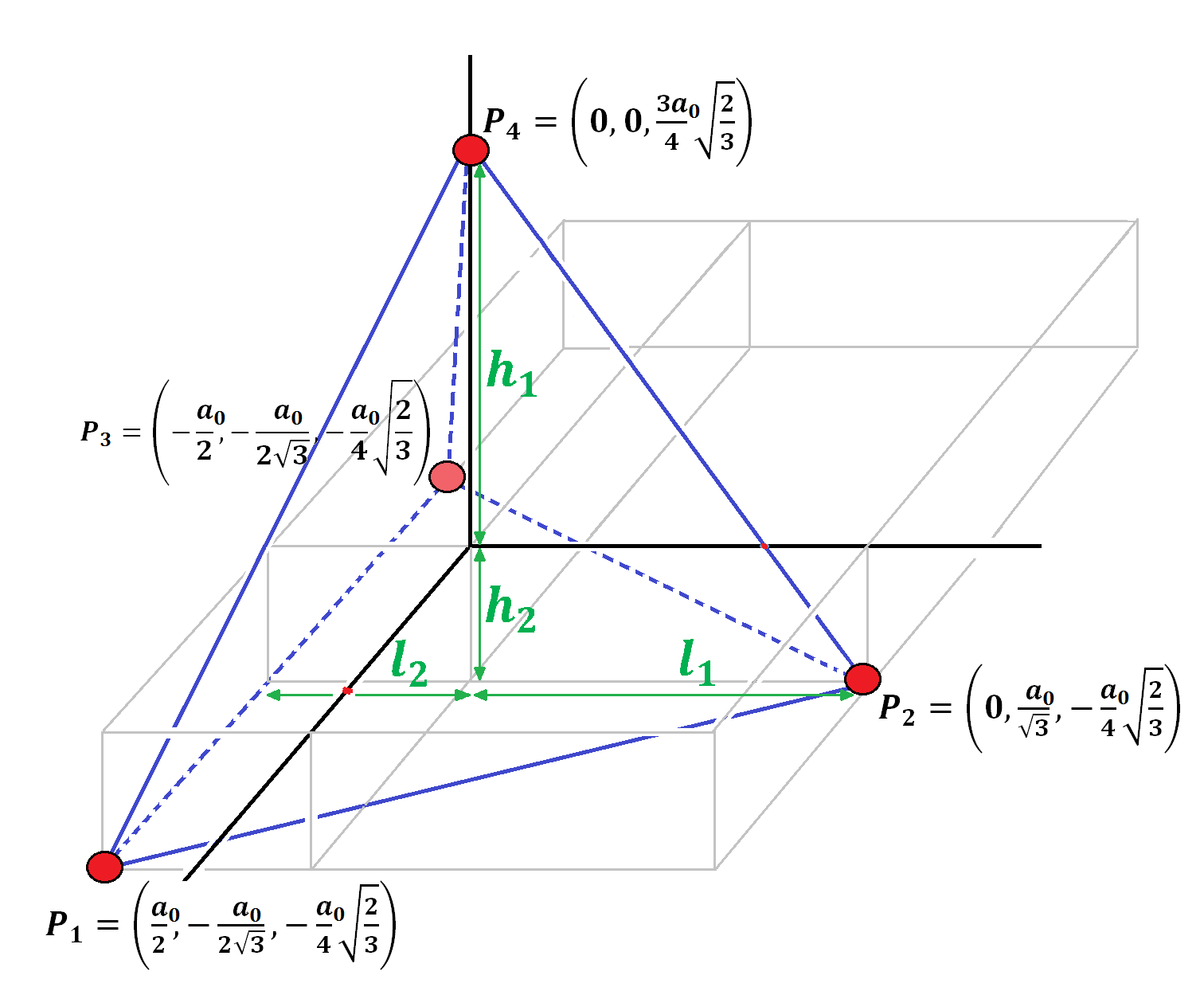}
\end{center}
\caption{A regular tetrahedron $T(a_0)$ in $\mathbb{R}^{3}$
(CKHcs).\label{fig2}}
\end{figure}
 \FloatBarrier

The ignorance of how this new coordinates system CKHcs can map an
Euclidean length to spherical and hyperbolic length measures, one
has to be careful in choosing the location of the tetrahedron
$T(a)$. From our choice in Fig. \ref{fig2}, it obvious to see that
the image of a regular geodesic tetrahedron $T(a)$ of an edge length
$a$ in the manifold is an Euclidean regular tetrahedron $T(a_0)$ of
a different edge length $a_0$ in the CKHcs:
\begin{equation}x\left(T(a)\right)=T(a_0)\,,\label{eq34}\end{equation}

Our objective is to have an expression for the starting Euclidean
length $a_0$ in terms of the geodesic length $a$. In order to
determine how this coordinates system measure the length different
from the original one, we have to consider two points
$M_1\left(x_1,y_1,z_1\right)$ and $M_2\left(x_2,y_2,z_2\right)$ in
the CKHcs where the corresponding  geodesic line between them is
parameterized by:
\begin{equation}\left\{ \begin{array}{l}
y=\alpha x+\beta  \\
z=\gamma x+\delta  \end{array} \right.\\,\label{eq35}\end{equation}
where
\begin{equation}\alpha =\frac{y_2-y_1}{x_2-x_1}\ \ \ \ \beta =\frac{x_2y_1-x_1y_2}{x_2-x_1}\,,\label{eq36}\end{equation}
\begin{equation}\gamma =\frac{z_2-z_1}{x_2-x_1}\ \ \ \ \delta =\frac{x_2z_1-x_1z_2}{x_2-x_1}\,,\label{eq37}\end{equation}

The geodesic length between $M_1$ and $M_2$ is:
\begin{equation}d\left(M_1M_2\right)=\epsilon \ r\ {\left.arctan\left(\frac{\left({\alpha }^2+{\gamma }^2+1\right)x+\alpha \beta +\gamma \delta }{\sqrt{{\epsilon }^2r^2+{\beta }^2+{\delta }^2+\left({\alpha }^2+{\gamma }^2\right){\epsilon }^2r^2+{\alpha }^2{\delta }^2+{\gamma }^2{\beta }^2-2\alpha \beta \gamma \delta }}\right)\right|}^{x_2}_{x_1}\,,\label{eq38}\end{equation}
Since $d\left(M_1M_2\right)$ depends strongly on the ending points,
a special care has to be done in the location of the Euclidean
regular tetrahedron in the CKHcs as it is shown in Fig. \ref{fig2}.
One can check that:
\begin{equation}a=2\epsilon r\ arctan\left(\frac{1}{2}\frac{a_0}{\sqrt{{\epsilon }^2r^2+\frac{a^2_0}{8}}}\right)\,,\label{eq39}\end{equation}
In order to obtain a geodesic edge length $a$, one has to solve Eq.
\eqref{eq39} for the unknown $a_0$ and get:
\begin{equation}a_0=\frac{2\ \epsilon \ r\ tan\left(\frac{a}{2\epsilon r}\right)}{\sqrt{\left(1-\frac{1}{2}{tan}^2\left(\frac{a}{2\epsilon r}\right)\right)}}\,,\label{eq40}\end{equation}
\begin{enumerate}
\item For the spherical case $S^3_r \ (\epsilon =1\Rightarrow R=\frac{6}{r^2})$ , one has:
\begin{equation}a=2r\ arctan\left(\frac{1}{2}\frac{a_0}{\sqrt{r^2+\frac{a^2_0}{8}}}\right)\,,\label{eq41}\end{equation}
In this case, one can check that the regular tetrahedron has a
maximal edge $a_{max}$ (for $a_0\to \infty $) given by:
\begin{equation}a_{max}=2\ arctan\left(\sqrt{2}\right)\ r\,,\label{eq42}\end{equation}

\item For the hyperbolic case $S^3_r \ (\epsilon =i\Rightarrow R=\frac{-6}{r^2})$ , one
has:
\begin{equation}a=2r\ arctanh\left(\frac{1}{2}\frac{a_0}{\sqrt{r^2-\frac{a^2_0}{8}}}\right)\,,\label{eq43}\end{equation}
Due to the compactness property (see Eq. \eqref{eq18}) of the
coordinates chart, the initial value of the Euclidean length $a_0$
must be bounded $a_0<\frac{2}{3}\sqrt{6}\ r$ . However, $a$ has no
upper bound.
\end{enumerate}

\subsection{Boundary area of a regular tetrahedron in $S^3_r$ and
$H^3_r$}

The faces area of a geodesic regular tetrahedron of an edge length
$a$ are all equal  $\left(A^{\mathrm{\Sigma
}}_f\left(r,a\right)=A^{\mathrm{\Sigma }}\left(r,a\right)\ , \
\forall f=\overline{1.4}\right)$ . In fact, the geodesic surface of
the $S^3_r$ and $H^3_r$ are portions of the great 2-dimensional
spheres $S^2_r$ and hyperbolic $H^2_r$ respectively. Accordingly, we
expect to obtain the same area expression of the spherical and
hyperbolic trigonometry. Due to the symmetric property of the
constant curvature spaces, we restrict ourselves to geodesic
triangle face ${\partial T(a)}_f \equiv P_{1}P_{2}P_{3}$ (See Fig.
\ref{fig2}) in the geodesic surface
$z=\frac{-a_{0}}{4}\sqrt{\frac{2}{3}}$ (with $dz=0$). Then the
induced Jacobian:
\begin{equation}\sqrt{|det(g(x)|_{P_{1}P_{2}P_{3}})|}=\frac{{\epsilon}^2 r^2 \sqrt{{\epsilon}^2 r^2+\frac{{a_0}^2}{24}}}{\left({\epsilon}^2r^2+x^2+y^2+\frac{{a_0}^2}{24}\right)^{3/2}},\end{equation}
The boundary face area is:
\begin{equation}A^{\mathrm{\Sigma }}\left(r,a\right)=\int_{P_{1}P_{2}P_{3}\subset {\mathbb{R}}^3}{{dA}^{Euc}_f}\frac{{\epsilon}^2 r^2 \sqrt{{\epsilon}^2r^2+\frac{{a_0}^2}{24}}}{\left({\epsilon }^2r^2+x^2+y^2+\frac{{a_0}^2}{24}\right)^{3/2}}\,,\label{eq44}\end{equation}
with
\begin{equation}{dA}^{Euc}_f=\frac{1}{2}\sum^3_{i,j,k=1}{{\epsilon }_{ijk}A^i_fdx^j\wedge dx^k}\,,\label{eq45}\end{equation}
where $A^i_f$ is the $i^{th}$ component of the normal area vector
${\vec{A}}_f$. The integral in Eq. \eqref{eq44} is in general very
hard to evaluate. To do so, one has to make a series expansion of
the Jacobian $J(\vec{x})$ given in \eqref{eq27} with respect to the
coordinates variables $\left\{\vec{x}\right\}\ $ and then easily
perform the integration over one of the faces $P_{1}P_{2}P_{3}$, we
get the following expression:
\begin{multline}A^{\mathrm{\Sigma }}(r,a)=\frac{\sqrt{3}}{4}a^2\{1+\frac{1}{8}{(\frac{a}{\epsilon r})}^2+\frac{1}{60}{(\frac{a}{\epsilon r})}^4+\frac{583}{241920}{(\frac{a}{\epsilon r})}^6\\+\frac{227}{604800}{(\frac{a}{\epsilon r})}^8+\frac{23}{369600}{(\frac{a}{\epsilon r})}^{10}+\frac{1418693}{130767436800}{(\frac{a}{\epsilon r})}^{12}+\mathcal{O}({(\frac{a}{\epsilon r})}^{14})\}\,,\label{eq46}\end{multline}
Using the symmetry of the triangle faces of a regular tetrahedron,
the exact formula of the boundary face area reads:
\begin{equation}A^{\mathrm{\Sigma }}\left(r,a\left(a_0\right)\right)=2\int^{\frac{a_0}{2}}_0{dx\int^{-\sqrt{3}x+\frac{\sqrt{3}a_0}{3}}_{\frac{-\sqrt{3}a_0}{6}}{dy}}\ \frac{{\epsilon}^2 r^2 \sqrt{{\epsilon }^2 r^2+\frac{{a_0}^2}{24}}}{\left({\epsilon}^2r^2+x^2+y^2+\frac{{a_0}^2}{24}\right)^{3/2}}\,,\label{eq47}\end{equation}
Straightforward but tedious calculations (See Appendix \ref{AppB})
give the following analytical expression of the boundary face area
$A^{\mathrm{\Sigma }}\left(r,a\right)$ of a regular spherical and
hyperbolic tetrahedron with an edge length $a$ in the curved space
$\mathrm{\Sigma }$ of a constant curvature $R=\frac{6}{{\epsilon
}^2r^2}$:
\begin{equation}A^{\mathrm{\Sigma }}\left(r,a\right)={\epsilon}^2 r^2\left(3\arccos\left(\frac{\cos(\frac{a}{\epsilon r})}{\cos(\frac{a}{\epsilon r})+1}\right)-\pi\right)\,,\label{eq48}\end{equation}
It is easy to check that the expansion of the resulted formula
\eqref{eq48} in terms of the $\frac{a}{\epsilon r}$ variable is
exactly the one in Eq. \eqref{eq46} and thus ensuring the
correctness of the integration.

\begin{enumerate}
\item For the spherical case $S^3_r \ (\epsilon =1\Rightarrow
R=\frac{6}{r^2})$ , one has:
\begin{equation}A^{S^3_r}\left(r,a\right)=r^2\left(3\arccos\left(\frac{\cos(\frac{a}{r})}{\cos(\frac{a}{r})+1}\right)-\pi\right)\,,\label{eq49}\end{equation}
As it is expected, it is the familiar expression of the regular
spherical triangle embedded in the 2-sphere $S^2_r$ where the
dihedral angle is defined by
$\Theta=\arccos\left(\frac{\cos(\frac{a}{r})}{\cos(\frac{a}{r})+1}\right)$
which is the cosine rule formula for spherical trigonometry. We can
check that the boundary area $A^{S^3_r}$ for the maximal edge length
$a_{max}$ in Eq. \eqref{eq42} corresponds to an upper bound
$A^{S^3_r}_{max}=\pi r^2$. The boundary area of a regular spherical
tetrahedron is always greater than the boundary area of a regular
Euclidean one.

\item For the hyperbolic case $S^3_r \ (\epsilon =i\Rightarrow
R=\frac{-6}{r^2})$ , one has:
\begin{equation}A^{H^3_r}\left(r,a\right)=r^2\left(\pi-3\arccos\left(\frac{\cosh(\frac{a}{r})}{\cosh(\frac{a}{r})+1}\right)\right)\,,\label{eq50}\end{equation}
As it is expected, it is the familiar expression of the regular
hyperbolic triangle embedded in the 2-hyperbolic $H^2_r$ where the
dihedral angle is defined by
$\Theta=\arccos\left(\frac{\cosh(\frac{a}{r})}{\cosh(\frac{a}{r})+1}\right)$
which is the cosine rule formula for hyperbolic trigonometry. Notice
that in this case, there is no upper bound and for a given pair
$(r,a)$. The boundary area of a regular hyperbolic tetrahedron is
always smaller than the boundary area of a regular Euclidean one.

\item For the Euclidean case $Euc^3 \ (R=0)$ , one has:
\begin{equation}A^{{Euc}^3}\left(r,a\right)={\mathop{lim}_{r\to \infty } A^{\mathit{\Sigma}}\left(r,a\right)\ }=\frac{\sqrt{3}}{4}a^2\,,\label{eq51}\end{equation}
The Euclidean limit is well-defined.
\end{enumerate}
\FloatBarrier
\begin{figure}[ht]
\begin{center}
\includegraphics[width=2.5in]{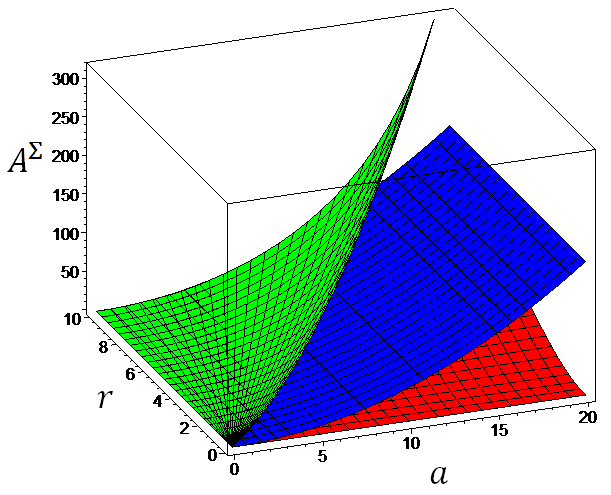}
\end{center}
\caption{: Function surface of the boundary face area for spherical
(green), Euclidean (blue) and hyperbolic (red) regular
tetrahedra.\label{fig3}}
\end{figure}
\FloatBarrier

\subsection{Volume of a regular tetrahedron in $S^3_r$ and
$H^3_r$}

The volume  $V^{\mathrm{\Sigma }}$ of a regular spherical and
hyperbolic tetrahedron is:
\begin{equation}V^{\mathrm{\Sigma }}\left(r,a\left(a_0\right)\right)=\int_{T\left(a_0\right)\subset {\mathbb{R}}^3}{{dV}^{Euc}\frac{r^4\ }{{\left({\epsilon }^2r^2+{\left|\vec{x}\right|}^2\right)}^2}}\,,\label{eq52}\end{equation}
Since the integration is very hard to deal with, it is better to
make again a series expansion of the Jacobian $J(\vec{x})$ given in
\eqref{eq27} in terms of the coordinates variables
$\left\{\vec{x}\right\}$ and then easily perform the integration to
end up with:
\begin{multline}V^{\mathrm{\Sigma }}(r,a)=\frac{\sqrt{2}}{12}a^3\{1+\frac{23}{80}{(\frac{a}{\epsilon r})}^2+\frac{3727}{53760}{(\frac{a}{\epsilon r})}^4+\frac{124627}{7741440}{(\frac{a}{\epsilon r})}^6 \\ +\frac{20283401}{5449973760}{(\frac{a}{\epsilon r})}^8+\frac{14700653069}{17003918131200}{(\frac{a}{\epsilon r})}^{10}+\frac{1651049434189}{8161880702976000}{(\frac{a}{\epsilon r})}^{12}+\mathcal{O}({\frac{a}{r})}^{14})\}\,,\label{eq53}\end{multline}
Using the symmetry of the regular tetrahedron, the exact expression
of the volume of a regular spherical and hyperbolic tetrahedron is:
\begin{equation}V^{\mathrm{\Sigma }}\left(r,a\left(a_0\right)\right)=2 \int^{\frac{\sqrt{6}a_0}{4}}_{\frac{-\sqrt{6}a_0}{12}}{dz\ \int^{\frac{\alpha(z)}{2}}_{0}{dx\ \int^{-\sqrt{3}x+\frac{\sqrt{3}\alpha(z)}{3}}_{\frac{-\sqrt{3}\alpha(z)}{6}}{dy\ \frac{r^4}{{\left({\epsilon }^2r^2+{\left|\vec{x}\right|}^2\right)}^2}}}}\,,\label{eq54}\end{equation}
where
\begin{equation}\alpha(z)=\frac{-\sqrt{6}}{2}z+\frac{3a_0}{4}\end{equation}
 Which can be rewritten in the following integral form (See
Appendix \ref{AppC}) as:
\begin{equation}V^{\mathrm{\Sigma }}\left(r,a\right)=12{\epsilon }^3\ r^3\int^{tan\left(\frac{a}{2\epsilon r}\right)}_0{dt\ \frac{\ t\ arctan\left(t\right)}{\left(3-t^2\right)\sqrt{2-t^2}}}\,,\label{eq55}\end{equation}
Notice that this integral has no analytic formula (we can carry the
integration by using numerical methods) and can be expressed in
terms of some special functions like the dilogarithm ${Li}_2(z)$,
the Clausen of order 2 ${Cl}_2\left(\varphi \right)$ or the digamma
$\mathrm{\Psi }\left(x\right)$. It is easy to check that the
expansion of the resulted formula \eqref{eq55} in terms of the
$\frac{a}{\epsilon r}$ variable is exactly the one in Eq.
\eqref{eq53} and thus ensuring the correctness of the integration.
\begin{enumerate}
\item For the spherical case $S^3_r \ (\epsilon =1\Rightarrow
R=\frac{6}{r^2})$ , one has:
\begin{equation}V^{S^3_r}\left(r,a\right)=12\ r^3\int^{tan\left(\frac{a}{2r}\right)}_0{dt\ \frac{\ t\ arctan\left(t\right)}{\left(3-t^2\right)\sqrt{2-t^2}}}\,,\label{eq56}\end{equation}
The volume for a maximal edge length
$V^{S^3_r}\left(r,a_{max}\right)$ (as it is expected)\textit{ }is
half of the 3-dimensional cubic hyperarea of 3-sphere of radius  $r$
:$\ $
\begin{equation}V^{S^3_r}\left(r,a_{max}\right)={\pi }^2r^3=\frac{1}{2}Area\left(S^3_r\subset {\mathbb{R}}^4\right)\,,\label{eq57}\end{equation}
Notice that for a given pair $(r,a)$ the volume of a regular
spherical tetrahedron is always greater than the regular Euclidean
one.

\item For the hyperbolic case $S^3_r \ (\epsilon =i\Rightarrow
R=\frac{-6}{r^2})$ , one has:
\begin{equation}V^{H^3_r}\left(r,a\right)=12\ r^3\int^{tanh\left(\frac{a}{2r}\right)}_0{dt\ \frac{\ t\ arctanh\left(t\right)}{\left(3+t^2\right)\sqrt{2+t^2}}}\,,\label{eq58}\end{equation}
has an upper bound :
\begin{equation}{\mathop{\mathrm{lim}}_{a\to \infty } V^{H^3_r}\left(r,a\right)\ }=1.0149416064096536250\ r^3\,,\label{eq59}\end{equation}
\begin{equation}\ =Im\left[{Li}_2\left(e^{i\frac{\pi }{3}}\right)\right]r^3=\frac{\sqrt{6}}{3}\left({\mathit{\Psi}}^1\left(\frac{1}{3}\right)-\frac{2}{3}{\pi }^2\right)r^3\ ={Cl}_2\left(\frac{\pi }{3}\right)r^3\ \,,\label{eq60}\end{equation}
Notice that for a given pair $(r,a)$ the volume of a regular
hyperbolic tetrahedron is always smaller than the regular Euclidean
one.

\item For the Euclidean case $Euc^3 \ (R=0)$ , one has:
\begin{equation}V^{{Euc}^3}\left(r,a\right)={\mathop{lim}_{r\to \infty } V^{\mathit{\Sigma}}\left(r,a\right)\ }=\frac{\sqrt{2}}{12}a^3\,,\label{eq61}\end{equation}
The Euclidean limit is well-defined.
\end{enumerate}

\FloatBarrier
\begin{figure}[ht]
\begin{center}
\includegraphics[width=2.5in]{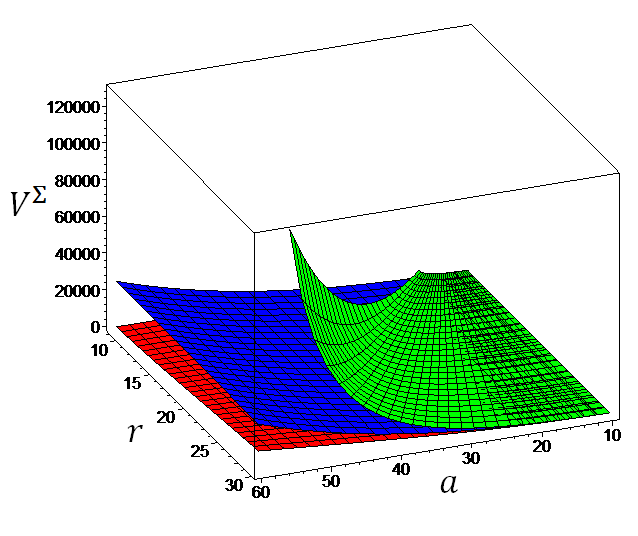}
\end{center}
\caption{Function surface of regular tetrahedron volume for
spherical (green), Euclidean (blue) and hyperbolic (red)
cases.\label{fig4}}
\end{figure}
 \FloatBarrier

\subsection{The volume-area ratio function}

We define the volume-area ratio function ${VRA}^{\mathrm{\Sigma }}$
for a regular geodesic tetrahedron as:
\begin{equation}{VRA}^{\mathit{\Sigma}}(r,a)=\frac{V^{\mathit{\Sigma}}\left(r,a\right)}{{\left(A^{\mathit{\Sigma}}\left(r,a\right)\right)}^{\frac{3}{2}}}\,,\label{eq62}\end{equation}
It is obvious that the ${VRA}^{\mathrm{\Sigma }}$ for a regular
Euclidean tetrahedron is a constant:
\begin{equation}{VRA}^{{Euc}^3}={\mathop{lim}_{r\to \infty } VRA(r,a)\ }=\frac{\sqrt{2}}{12{\left(\frac{\sqrt{3}}{4}\right)}^{\frac{3}{2}}}=0.4136\,,\label{eq63}\end{equation}

\begin{figure}[ht]
\begin{center}
\includegraphics[width=2.5in]{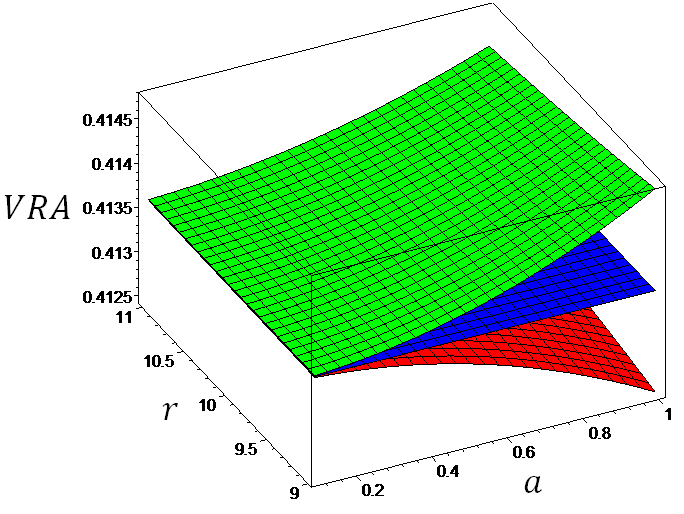}
\end{center}
\caption{The volume-area ratio function for spherical (green),
Euclidean (blue) and hyperbolic (red) cases.\label{fig5}}
\end{figure}

\begin{corollary}
according to the useful inequality
\begin{equation}{VRA}^{H^3_r}(r,a)\le {VRA}^{{Euc}^3}(r,a)\le {VRA}^{S^3_r}(r,a)\,,\label{eq64}\end{equation}
the  ${VRA}^{\mathrm{\Sigma }}$ function allows us to know what kind
of geometry inside the regular geodesic tetrahedron: (see Fig.
\ref{fig5})
\begin{equation}\left\{ \begin{array}{l}
{VRA}^{\mathit{\Sigma}}(r,a)> 0.4136\ \ \ \ \ \ \ \ \ \ S^3_r\\
{VRA}^{\mathit{\Sigma}}(r,a)= 0.4136\ \ \ \ \ \ \ \ \ \ Euc^3\\
 {VRA}^{\mathit{\Sigma}}\left(r,a\right)< 0.4136\ \ \ \ \ \ \ \ \ \ H^3_r\end{array}
\right.\,,\label{eq65}\end{equation}
\end{corollary}

\FloatBarrier
 \FloatBarrier

\subsection{The volume function in terms of scalar curvature and area}

\begin{figure}[ht]
\begin{center}
\includegraphics[width=5in]{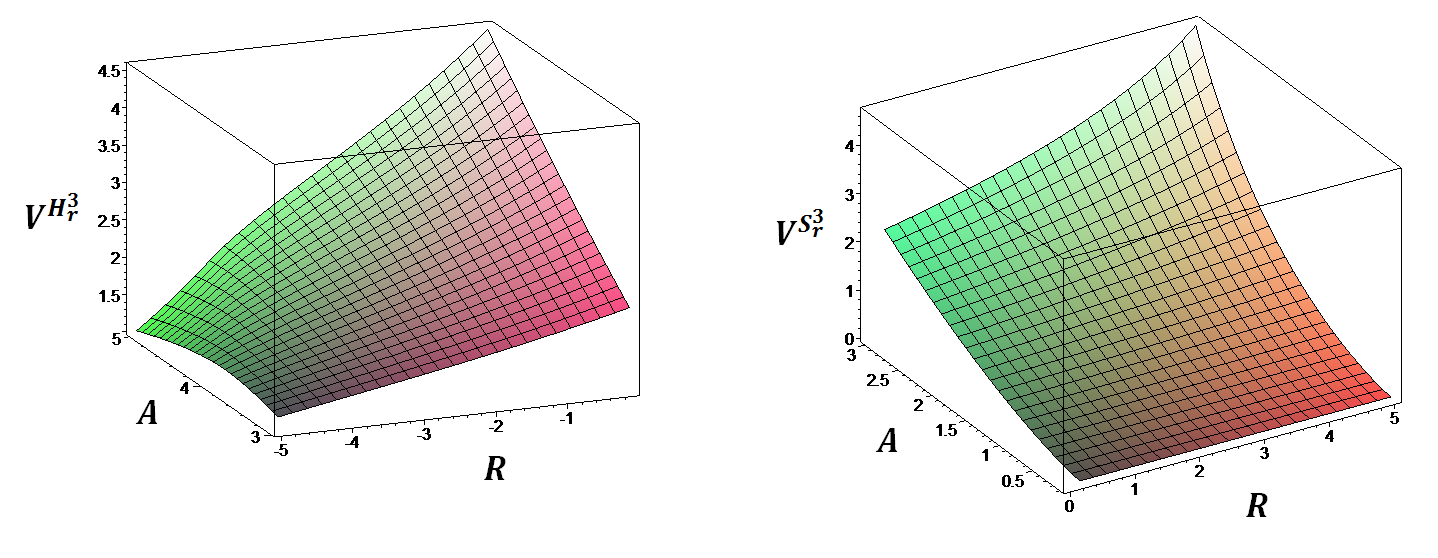}
\end{center}
\caption{The volume function in terms of scalar curvature $R$ and
area $A$ for spherical (right) and hyperbolic (left) regular
tetrahedron.\label{fig6}}
\end{figure}

From the area formula \eqref{eq48}, one can express the edge length
$a$ by:
\begin{equation}a(A,R)=\left(\pi-\arccos\left(\frac{\sin(\frac{-\pi}{6}+\frac{A}{3{\epsilon}^2{r}^2})}{\sin(\frac{-\pi}{6}+\frac{A}{3{\epsilon}^2{r}^2})+1}\right)\right)\epsilon r\,,\label{eq66}\end{equation}
substitute it in Eq. \eqref{eq55} to get a volume function in terms
of the 3d- Ricci scalar curvature and boundary face area of a
regular tetrahedron:
\begin{equation}V^{\mathrm{\Sigma }}=V^{\mathrm{\Sigma }}\left(R,a\left(R,A\right)\right)=V^{\mathrm{\Sigma }}\left(R,A\right)\,,\label{eq68}\end{equation}
\begin{corollary} the volume of a regular geodesic tetrahedron for a fixed boundary area satisfies the following inequality
\begin{equation}For \ any \ \ \ R_1,R_2\in \mathbb{R} \ \ \ if \ \ \ R_1<R_2 \ \ \ then \ \ \
V^{\mathrm{\Sigma }}(R_1,A)<V^{\mathrm{\Sigma
}}(R_2,A)\,,\label{eq69}\end{equation}

this results from the fact that the function $V^{\mathrm{\Sigma }}$
increases with respect to $R$ for a fixed area norm $A$ (see Fig.
\ref{fig6}).

\end{corollary}

\FloatBarrier
 \FloatBarrier

\section{Conclusion}\label{Sec5}

In this paper, we explicitly derived the boundary face area and
volume of a regular spherical and hyperbolic tetrahedron in terms of
the curvature radius (or the scalar curvature) and the edge length.
We have directly performed the integration over the area and volume
elements by using the Cayley-Klein-Hilbert coordinates system
(CKHcs) to end up with simple formula given in Eqs.
(\ref{eq48},\ref{eq55}). A comparison between the Euclidean,
spherical and hyperbolic cases is studied and their implications are
discussed. It is shown that the volume function of a regular
geodesic tetrahedron for a fixed boundary face area is a strictly
increasing in the scalar curvature interval.
\appendix
\section{Proof of the relation \eqref{eq2}}\label{AppA}
The geodesics in the CKHcs $\{U\subset M,\vec{x}\}$ are straight
lines, one has:
\begin{equation}\ddot{x}^A=0\,,\label{eqA1}\end{equation}
The condition
\begin{equation}\Gamma^{A}_{BC}(x)
\dot{x}^B\dot{x}^C=0\,,\label{eqA2}\end{equation} must be hold,
which implies:
\begin{equation}\Gamma^{A}_{BC}(x) \frac{{\partial \varphi}^B(\tilde{x})}{{\partial\tilde{x}}^I}\frac{{\partial \varphi}^C(\tilde{x})}{{\partial\tilde{x}}^J}{\dot{\tilde{x}}}^I{\dot{\tilde{x}}}^J=0\,,\label{eqA3}\end{equation}
Under the transformation \eqref{eq1}, the Christoffel symbols
transform as:
\begin{equation}\Gamma^{A}_{BC}(x)=\frac{{\partial \tilde{x}}^J}{{\partial x}^B}\frac{{\partial \tilde{x}}^K}{{\partial x}^C}\frac{{\partial \varphi}^A}{{\partial
\tilde{x}}^I}{\tilde{\Gamma}}^{I}_{JK}(\tilde{x})-\frac{{\partial
\tilde{x}}^J}{{\partial x}^B}\frac{{\partial \tilde{x}}^K}{{\partial
x}^C}\frac{{\partial^2 \varphi}^A}{{\partial \tilde{x}}^J{\partial
\tilde{x}}^K}\,,\label{eqA4}\end{equation} By substituting it in Eq.
\eqref{eqA3}, one can obtain the transformation condition Eq.
\eqref{eq2} to the ideal CKHcs frame.

\section{Proof of the area formula}\label{AppB}
The boundary face area ($P_{1}P_{2}P_{3}$) of a regular spherical
and hyperbolic tetrahedron of an edge length $a$ is given by an
integral form in Eq. \eqref{eq47}. For simplicity, we drop the
triangle face $P_{1}P_{2}P_{3}$ to $\Pi(P_{1}P_{2}P_{3})$ in the
$XY$-plane (since the area of a fixed triangle is the same wherever
its location inside the constant curvature manifold). In this case,
the induced Jacobian can be written as:
\begin{equation}\sqrt{|det(g(x)|_{\Pi(P_{1}P_{2}P_{3})})|}=\frac{{\epsilon}^3 r^3}{({\epsilon}^2r^2+x^2+y^2)^{3/2}},\label{B1}\end{equation}
The boundary face area is given by:
\begin{equation}A^{\mathrm{\Sigma }}\left(r,a\right)=2\int^{\frac{a_0}{2}}_0{dx\int^{-\sqrt{3}x+\frac{\sqrt{3}a_0}{3}}_{\frac{-\sqrt{3}a_0}{6}}{dy}}\ \frac{{\epsilon}^3 r^3}{\left({\epsilon }^2r^2+x^2+y^2\right)^{3/2}},\label{B2}\end{equation}
where one can check the starting Euclidean length $a_0$ in this case
is given by:
\begin{equation}a_0=2\epsilon r \frac{\tan(\frac{a}{2\epsilon r})}{\sqrt{1-\frac{1}{3}\tan(\frac{a}{2\epsilon r})^2}},\label{B3}\end{equation}
Performing the Integral over y variable, one get:
\begin{multline}\int^{-\sqrt{3}x+\frac{\sqrt{3}a_0}{3}}_{\frac{-\sqrt{3}a_0}{6}}{dy}\ \frac{{\epsilon}^3 r^3}{\left({\epsilon }^2r^2+x^2+y^2\right)^{3/2}}=\\\frac{{{\epsilon}^3}r^3(-\sqrt{3}x+\frac{\sqrt{3}a_{0}}{3})}{({{\epsilon}^2}r^2+x^2)\sqrt{{{\epsilon}^2}r^2+x^2+(-\sqrt{3}x+\frac{\sqrt{3}a_{0}}{3})^2}}+\frac{{{\epsilon}^3}r^3\frac{\sqrt{3}a_0}{6}}{({{\epsilon}^2}r^2+x^2)\sqrt{{{\epsilon}^2}r^2+x^2+\frac{a_{0}^{2}}{12}}},\label{B4}\end{multline}
Let us preform the second integral over the x variable. By
integrating each term separately, one has:
\begin{multline}t_1(x)=\int^{\frac{a_0}{2}}_0{dx \frac{{{\epsilon}^3}r^3(-\sqrt{3}x+\frac{\sqrt{3}a_{0}}{3})}{({{\epsilon}^2}r^2+x^2)\sqrt{{{\epsilon}^2}r^2+x^2+(-\sqrt{3}x+\frac{\sqrt{3}a_{0}}{3})^2}}}={\epsilon}^2r^2\arctan(\frac{F(a_0,r;x)}{G(a_0,r;x)}),\label{B5}\end{multline}
where
\begin{multline}F(a_0,r;x)=-\frac{\sqrt{3}}{3}\sqrt{{\epsilon}^2r^2+4x^2-2xa_0+\frac{a_{0}^{2}}{3}}({\epsilon}^2r^2+\frac{a_{0}^{2}}{9})(-{\epsilon}^2r^2+\frac{a_{0}x}{3})\\-a_{0}(\frac{5{\epsilon}^2r^2}{9}+\frac{a_{0}^{2}}{27})({\epsilon}^2r^2+x^2)+\frac{a_{0}^{4}x}{81}+r^4x+\frac{2a_{0}^{2}{\epsilon}^2r^2x}{9},\label{B6}\end{multline}
and
\begin{multline}G(a_0,r;x)=\frac{\epsilon
r\sqrt{3}}{3}\sqrt{{\epsilon}^2r^2+4x^2-2xa_0+\frac{a_{0}^{2}}{3}}({\epsilon}^2r^2+\frac{a_{0}^{2}}{9})(x+\frac{a_{0}}{3})\\+\frac{2a_{0}^{2}\epsilon
rx^2}{27}+\frac{{\epsilon}^5r^5}{3}-\frac{4a_{0}^{2}{\epsilon}^3r^3}{27}-\frac{a_{0}^{4}{\epsilon}r}{81}+\frac{4{\epsilon}^3r^3x^2}{3},\label{B7}\end{multline}
\begin{equation}t_2(x)=\int^{\frac{a_0}{2}}_0{dx \frac{{{\epsilon}^3}r^3\frac{\sqrt{3}a_0}{6}}{({{\epsilon}^2}r^2+x^2)\sqrt{{{\epsilon}^2}r^2+x^2+\frac{a_{0}^{2}}{12}}}}={\epsilon}^2r^2\arctan\left(\frac{a_{0}x}{\epsilon r\sqrt{12{{\epsilon}^2}r^2+12x^2+a_{0}^{2}}}\right),\label{B8}\end{equation}
Adding the two terms together, we obtain:
\begin{multline}A^{\mathrm{\Sigma }}\left(r,a\right)=2(t_{1}(x)+t_{2}(x))|^{x=a_0/2}_{x=0}=\\2{\epsilon}^2r^2\arctan(\frac{9a_{0}^{2}{\epsilon}r(3a_{0}^{2}-\sqrt{3}a_{0}\sqrt{9{\epsilon}^2r^2+3a_{0}^{2}}+18{\epsilon}^2r^2)}{3a_{0}^{5}-63a_{0}^{3}{\epsilon}^2r^2-216a_{0}r^4+\sqrt{3}\sqrt{9{\epsilon}^2r^2+3a_{0}^{2}}(18a_{0}^{2}{\epsilon}^2r^2+144r^4-a_{0}^{4})})\label{B9}\end{multline}
When we replace $a_0$ given in Eq. \eqref{B3}, we get the area
function formula of Eq. \eqref{eq48}.
\section{Proof of the volume formula}\label{AppC}
The volume of a regular spherical and hyperbolic tetrahedron of an
edge length $a$ is given by an integral form in Eq. \eqref{eq54}.
Using the integration by shell method (taking the sum of parallel
triangles of constant $z$). Performing the Integral over the $y$
variable, one get:
\begin{multline}\int^{-\sqrt{3}x+\frac{\sqrt{3}\alpha(z)}{3}}_{\frac{-\sqrt{3}\alpha(z)}{6}}{dy}\frac{r^4\
}{{\left({\epsilon
}^2r^2+x^2+y^2+\frac{{\alpha(z)}^2}{24}\right)}^2}=\\\frac{32\sqrt{3}\
r^4\ \left(-3x+\alpha(z)\right)}{\left(32x^2-16\alpha(z)x+8{\epsilon
}^2r^2+3{\alpha(z)}^2\right)\left(24x^2+24{\epsilon
}^2r^2+{\alpha(z)}^2\right)}+\frac{24\sqrt{6}\ r^4\
arctan\left(\frac{2\sqrt{2}\left(-3x+\alpha(z)\right)}{\sqrt{24x^2+24{\epsilon
}^2r^2+{\alpha(z)}^2}}\right)}{\left(24x^2+24{\epsilon
}^2r^2+{\alpha(z)}^2\right)}\\+\frac{48\sqrt{3}\ r^4\alpha(z)\
}{\left(24x^2+24{\epsilon
}^2r^2+3{\alpha(z)}^2\right)\left(24x^2+24{\epsilon
}^2r^2+{\alpha(z)}^2\right)}+\frac{24\sqrt{6}\ r^4\
arctan\left(\frac{\sqrt{2}\alpha(z)}{\sqrt{24x^2+24{\epsilon
}^2r^2+{\alpha(z)}^2}}\right)}{{\left(24x^2+24{\epsilon
}^2r^2+{\alpha(z)}^2\right)}^{\frac{3}{2}}}\,,\label{eqB1}\end{multline}
Now, let us focus on the second integral over the $x$ variable. By
integrating each term separately, one has:
\begin{multline}T_1\ (x)=\int{dx\frac{32\sqrt{3}\ r^4\ \left(-3x+\alpha(z)\right)}{\left(32x^2-16\alpha(z)x+8{\epsilon }^2r^2+3{\alpha(z)}^2\right)\left(24x^2+24{\epsilon }^2r^2+{\alpha(z)}^2\right)}}=\\ \frac{-6\sqrt{3}\ r^4\ ln\left(32x^2-16\alpha(z)x+8{\epsilon }^2r^2+3{\alpha(z)}^2\right)}{72{\epsilon }^2r^2+11{\alpha(z)}^2}-\frac{8\sqrt{3}\ r^4\alpha(z)arctan\left(\frac{8x-2\alpha(z)}{\sqrt{16{\epsilon }^2r^2+2{\alpha(z)}^2}}\right)}{\left(72{\epsilon }^2r^2+11{\alpha(z)}^2\right)\sqrt{16{\epsilon }^2r^2+2{\alpha(z)}^2}}\\+\frac{6\sqrt{3}\ r^4\ r^4\ ln\left(24x^2+24{\epsilon }^2r^2+{\alpha(z)}^2\right)}{72{\epsilon }^2r^2+11{\alpha(z)}^2}+\frac{48\sqrt{3}\ r^4\alpha(z)arctan\left(\frac{12x}{\sqrt{144{\epsilon }^2r^2+6{\alpha(z)}^2}}\right)}{\left(72{\epsilon }^2r^2+11{\alpha(z)}^2\right)\sqrt{144{\epsilon }^2r^2+6{\alpha(z)}^2}}\,,\label{eqB2}\end{multline}
\begin{multline}T_2\ (x)=\int{dx\frac{24\sqrt{6}\ r^4\ arctan\left(\frac{2\sqrt{2}\left(-3x+\alpha(z)\right)}{\sqrt{24x^2+24{\epsilon }^2r^2+{\alpha(z)}^2}}\right)}{\left(24x^2+24{\epsilon }^2r^2+{\alpha(z)}^2\right)}}=\\ \frac{48\sqrt{6}\ r^4\sqrt{8{\epsilon }^2r^2+{\alpha(z)}^2}\mathrm{\ arctan}\mathrm{}\left(\frac{\sqrt{2}\left(4x-\alpha(z)\right)}{\sqrt{8{\epsilon }^2r^2+{\alpha(z)}^2}}\right)}{\left(24{\epsilon }^2r^2+{\alpha(z)}^2\right)\left(72{\epsilon }^2r^2+11{\alpha(z)}^2\right)}-\frac{6\sqrt{3}\ r^4\ ln\left(24x^2+24{\epsilon }^2r^2+{\alpha(z)}^2\right)}{\left(72{\epsilon }^2r^2+11{\alpha(z)}^2\right)}\\ +\frac{6\sqrt{3}\ r^4\ ln\left(96x^2-48\alpha(z)x+24{\epsilon }^2r^2+9{\alpha(z)}^2\right)}{\left(72{\epsilon }^2r^2+11{\alpha(z)}^2\right)}-\frac{24\sqrt{2}\ arctan\left(\frac{\sqrt{24}\ x^2}{\sqrt{24{\epsilon }^2r^2+{\alpha(z)}^2}}\right)}{\left(72{\epsilon }^2r^2+11{\alpha(z)}^2\right)\sqrt{24{\epsilon }^2r^2+{\alpha(z)}^2}}\\ +\frac{24\sqrt{6}\ r^4\ x\ arctan\left(\frac{2\sqrt{2}\left(-3x+\alpha(z)\right)}{\sqrt{24x^2+24{\epsilon }^2r^2+{\alpha(z)}^2}}\right)}{\left(24{\epsilon }^2r^2+{\alpha(z)}^2\right)\sqrt{24x^2+24{\epsilon }^2r^2+{\alpha(z)}^2}}\,,\label{eqB3}\end{multline}
\begin{multline}T_3\ (x)=\int{dx\ \frac{48\sqrt{3}\ r^4\alpha(z)\ }{\left(24x^2+24{\epsilon }^2r^2+3{\alpha(z)}^2\right)\left(24x^2+24{\epsilon }^2r^2+{\alpha(z)}^2\right)}}=\\ \frac{6\sqrt{2}\ r^4\ arctan\left(\frac{2\sqrt{6}\ x}{\sqrt{24{\epsilon }^2r^2+{\alpha(z)}^2}}\right)}{\alpha(z)\sqrt{24{\epsilon }^2r^2+{\alpha(z)}^2}}-\frac{2\sqrt{6}\ r^4\ arctan\left(\frac{2\sqrt{2}\ x}{\sqrt{8{\epsilon }^2r^2+{\alpha(z)}^2}}\right)}{\alpha(z)\sqrt{8{\epsilon }^2r^2+{\alpha(z)}^2}}\,,\label{eqB4}\end{multline}
\begin{multline}T_4\ (x)=\int{dx\ \frac{24\sqrt{6}\ r^4\ arctan\left(\frac{\sqrt{2}\alpha(z)}{\sqrt{24x^2+24{\epsilon }^2r^2+{\alpha(z)}^2}}\right)}{{\left(24x^2+24{\epsilon }^2r^2+{\alpha(z)}^2\right)}^{\frac{3}{2}}}}=\\ \frac{\ 6\sqrt{6}\ r^4\sqrt{8{\epsilon }^2r^2+{\alpha(z)}^2}\ arctan\left(\frac{2\sqrt{2}\ x}{\sqrt{8{\epsilon }^2r^2+{\alpha(z)}^2}}\right)}{\left(24{\epsilon }^2r^2+{\alpha(z)}^2\right)}+\frac{24\sqrt{6}\ r^4x\ arctan\left(\frac{\sqrt{2}\alpha(z)}{\sqrt{24x^2+24{\epsilon }^2r^2+{\alpha(z)}^2}}\right)}{\left(24{\epsilon }^2r^2+{\alpha(z)}^2\right)\sqrt{24x^2+24{\epsilon }^2r^2+{\alpha(z)}^2}}\\-\frac{6\sqrt{2}\ r^4\ arctan\left(\frac{2\sqrt{6}\ x}{\sqrt{24{\epsilon }^2r^2+{\alpha(z)}^2}}\right)}{\alpha(z)\sqrt{24{\epsilon }^2r^2+{\alpha(z)}^2}}\,,\label{eqB5}\end{multline}
Adding all four terms together, we obtain:
\begin{equation}\begin{array}{c}
2{\left.\left(T_1\ \left(x\right)+T_2\ \left(x\right)+T_3\
\left(x\right)+T_4\
\left(x\right)\right)\right|}^{x=\alpha(z)/2}_{x=0}\\
=\frac{24\sqrt{6}\ r^4\ \alpha(z)\
arctan\left(\frac{\sqrt{2}\alpha(z)}{\sqrt{8{\epsilon
}^2r^2+{\alpha(z)}^2}}\right)}{\left(24{\epsilon
}^2r^2+{\alpha(z)}^2\right)\sqrt{8{\epsilon }^2r^2+{\alpha(z)}^2}}
 \end{array}
\,,\label{eqB6}\end{equation} Making the following change of
variable in the third integral over $z$:
\begin{equation}t=\frac{\sqrt{2}\alpha(z)}{\sqrt{8\epsilon r^2+{\alpha(z)}^2}}\,,\label{eqC1}\end{equation}
When we replace $a_0$ given in Eq. \eqref{eq40}, we get the volume
function formula of Eq. \eqref{eq55}.

\FloatBarrier


\begin{thebibliography}{0}

\bibitem{r1} D.V. Alekseevskii, E.B. Vinberg, A.S. Solodovnikov "Geometry-2:
Geometry of spaces of constant curvature," Encycl. of Math. Sci.
(Springer-Verlag), vol. 29, 1-146, (1993).
\bibitem{r2} G. J. Martin, The volume of regular tetrahedra and sphere packing in hyperbolic 3-space, Math. Chronicle 20,
127–147, (1991).
\bibitem{r3} Y. Cho and H. Kim, On the volume formula
for hyperbolic tetrahedra, Discrete Comput. Geom. 22, no. 3,
347–366, (1999).
\bibitem{r4} J. Murakami, M. Yano, On the volume of hyperbolic and spherical
tetrahedron, Comm. Annal. Geom. 13 (2), 379-400 (2005).
\bibitem{r5} D. A. Derevnin, A. D. Mednykh, A formula for the volume of a hyperbolic tetrahedon, Uspekhi Mat. Nauk, 60:2(362), 159–160; Russ. Math. Surveys,
346–348, (2005).
\bibitem{r6} Ushijima A, A Volume Formula for Generalised Hyperbolic Tetrahedra, In: Prékopa A., Molnár E. (eds) Non-Euclidean Geometries, Math. App, vol 581, Springer, Boston,
MA, (2006).
\href{https://arxiv.org/abs/math/0309216}{arXiv:math.GT/0309216}.
\bibitem{r7} J. Murakami, A. Ushijima, A volume formula for hyperbolic tetrahedra
in terms of edge lengths, J. Geom. 83 (1-2), 153-163, (2005),
\href{https://arxiv.org/abs/math/0402087}{arXiv:math.MG/0402087}.
\bibitem{r8} A. Kolpakov, A. Mednykh, M. Pashkevich, Volume formula for a Z2-symmetric spherical
tetrahedron through its edge lengths, Ark. Mat. Volume 51, Number 1,
99-123, (2013),
\href{https://arxiv.org/abs/1007.3948}{arXiv:math.MG/1007.3948}.
\bibitem{r9} Jun Murakami, Volume formulas for a spherical
tetrahedron, Proc. Amer. Math. Soc. 140, 3289-3295, (2012),
\href{https://arxiv.org/abs/1011.2584}{arXiv:math.MG/1011.2584}.

\bibitem{r10} E. Bianchi, P. Dona and S. Speziale, Polyhedra in loop quantum gravity, Phys. Rev. D 83, 044035
(2011),
\href{https://arxiv.org/abs/1009.3402v2}{arXiv:gr-qc/1009.3402}.
\bibitem{r11} O. Nemoul, N. Mebarki, A New Curvature Operator for a Regular Tetrahedron Shape in
LQG, (2018),
\href{https://arxiv.org/abs/1803.03134}{arXiv:gr-qc/1803.03134}.
\bibitem{r12} H. M. Haggard, Mu. Han, A. Riello, Encoding Curved Tetrahedra in Face Holonomies: a Phase Space of
Shapes from Group-Valued Moment Maps, Annales Henri Poincar\'e 17
(2016) no.8, 2001-2048,
\href{https://arxiv.org/abs/1506.03053}{arXiv:gr-qc/1506.03053}.
\bibitem{r13} Y. Taylor and C. Woodward, 6j symbols for Uq (sl2) and non-Euclidean tetrahedra, Sel. Math. New. Ser. 11 (2005),
539, \href{https://arxiv.org/abs/math/0305113}{arXiv:math.QA/0305113
}.
\bibitem{r14} M. Dupuis, F. Girelli, Observables in Loop Quantum Gravity with a cosmological
constant, Phys. Rev. D 90, 104037 (2014),
\href{https://arxiv.org/abs/1311.6841}{arXiv:gr-qc/1311.6841}.
\bibitem{r15} Barrett J W, First order Regge calculus Class. Quantum Grav. 11 2723, (1994), \href{https://arxiv.org/abs/hep-th/9404124}{arXiv:hep-th/9404124}.
\bibitem{r16} B. Bahr, B. Dittrich, Regge calculus from a new angle, New
J.Phys. 12:033010, (2010),
\href{https://arxiv.org/abs/0907.4325}{arXiv:gr-qc/0907.4325}.
\bibitem{r17} S. Ariwahjoedi, F. P. Zen, (2+1) Regge Calculus:
Discrete Curvatures, Bianchi Identity, and Gauss-Codazzi Equation,
(2017),
\href{https://arxiv.org/abs/1709.08373}{arXiv:gr-qc/1709.08373}.

\end{thebibliography}
\end{document}